\documentclass[final]{aipproc}

\layoutstyle{6x9}

\newcommand{\Teff}{\mbox{$T_\mathrm{eff}$}}

\newcommand{\lppr}{\stackrel{<}{\scriptstyle \sim}}
\newcommand{\lappr}{\raisebox{-0.4ex}{$\lppr $}}

\newcommand{\sn}{SN~1987A}
\newcommand{\msun}{\ensuremath{\, {\rm M}_\odot}}
\newcommand{\mdot}{\mbox{$\dot{M}$}}

\begin{document}

\title{Limits on iron-dominated fallback disk in SN~1987A}


\author{K. Werner}{
  address={Institute for Astronomy and Astrophysics, University
  of T\"ubingen, 72076 T\"ubingen, Germany}
}

\author{T. Nagel}{
  address={Institute for Astronomy and Astrophysics, University
  of T\"ubingen, 72076 T\"ubingen, Germany}
}

\author{T. Rauch}{
  address={Institute for Astronomy and Astrophysics, University
  of T\"ubingen, 72076 T\"ubingen, Germany}
}

\begin{abstract}
The non-detection of a point source in \sn\ imposes an upper limit
for the optical luminosity of $L_{\rm opt}\lappr 2$~L$_{\rm \odot}$.  This
limits the size of a possible fallback disk around the stellar remnant. Assuming
a steady-state thin disk with blackbody emission requires a disk smaller than
100\,000~km if the accretion rate is at 30\% of the Eddington rate
\citep{graves05}. We have performed detailed non-LTE radiation transfer
calculations to model the disk spectrum more realistically. It turns out that
the observational limit on the disk extension becomes even tighter, namely
70\,000~km.
\end{abstract}

\maketitle

\section{Introduction}

To date there is no direct evidence for a compact object at the center of the
\sn\ remnant. The initial neutrino burst is still the only evidence for
the formation of a compact object which could be either a neutron star or a
black hole. The non-detection of a point source in HST images imposes a tight
upper limit to its optical luminosity, which cannot exceed about 2~L$_{\rm
\odot}$ \citep{graves05}.

This allows to draw tight limits on the size of a possible supernova
fallback-disk around the compact object. Assuming steady-state thin-disk
accretion the optical/UV luminosity depends on the mass-accretion rate and outer
disk radius. For example, it was concluded that the observed optical flux limit
requires a small disk, no larger than 10$^5$~km, with an accretion rate of no more
than 30\% of the Eddington accretion rate \citep{graves05}. This result was
obtained assuming a steady-state Shakura-Sunyaev disk \citep{shakura73} composed
of concentric rings emitting blackbody spectra. It is our aim to see how the
limits on the fallback disk are affected by the blackbody assumption and to this
end we compute more realistic disk spectra by detailed
radiation-transfer calculations.

Aside from the particular case of \sn, observational proofs for the existence of
SN-fallback disks around pulsars are still debated. The X-ray
luminosities of anomalous X-ray pulsars (AXPs), which are slowly rotating
($P_{\rm rot}=5-12$~s), young ($\leq 100\,000$~yr), isolated neutron stars, are
generally assumed to be powered by magnetic energy \citep{woods06}.  As an
alternative explanation the X-ray emission was attributed to accretion from a
disk that is made up of SN-fallback material \citep{alp01,chat00,vanp95}.

The fallback-disk model, however, has difficulties to explain IR/optical
emission properties of AXPs.  When compared with disk models, the faint
IR/optical flux suggests that any disk around AXPs must be very compact (e.g.\
\citep{perna00,israel04}).  The discovery of optical pulsations in the AXP
4U\,0142+61 which have the same period like the X-ray pulsations \citep{kern02}
appears to be a strong argument against the disk model. It was argued that
reprocessing of the pulsed neutron star X-ray emission in the disk cannot explain the high
optical pulsed fraction, because disk radiation would be dominated by viscous
dissipation and not by reprocessed neutron star irradiation \citep{kern02}. On the other
hand, it was shown that these optical pulsations can  be explained either by the
magnetar outer gap model or by the disk-star dynamo  model
\citep{ertan04}. Therefore, the observation of optical pulsations is not an
argument  against the disk model.

The recent discovery of mid-IR emission from this AXP \citep{wang06}, however,
has strongly rekindled the interest in studies of fallback-disk emission
properties. While this mid-IR emission is attributed to a cool, passive (X-ray
irradiated) dust debris disk \citep{wang06}, all optical and near-IR flux
measurements can be  explained with a model for an active, dissipating gas disk
\citep{ertan07}.

Coming back to \sn, the presence of a fallback disk around its stellar remnant
has been invoked in order to explain its observed lightcurve which deviates from
the theoretical one for pure radioactive decay \citep{meyer92}.

\section{Model assumptions}

We employ our computer code {\sc AcDc} \citep{Na04}, that
calculates disk spectra under the following assumptions. The {\bf radial disk
structure} is calculated assuming a stationary, Keplerian, geometrically thin
$\alpha$-disk \citep{shakura73}. As pointed out in \citep{menou01}, for a
comparison with observational data one probably has to use a more elaborate
model, because near the outer disk edge the viscous dissipation and hence the
surface mass density decline stronger with increasing radius than in an
$\alpha$-disk. However, the purpose of the present paper is to look for
differential effects of various assumptions. Qualitatively, these effects can be
expected to be independent of the detailed radial disk structure. In any case,
it would be no problem to carry out the computations presented here with
different radial structures.

The $\alpha$-disk model is fixed by four global input parameters: Stellar mass
$M_\star$ and radius $R_\star$ of the accretor, mass accretion rate \mdot, and
the viscosity parameter $\alpha$. For numerical treatment the disk is
represented by a number of concentric rings. For each ring with radius $R$ our code
calculates the detailed vertical structure, assuming a plane-parallel radiating
slab.

A particular disk ring with radius $R$ is characterized by the following two
parameters, which follow from the global disk parameters introduced above. The
first parameter measures the dissipated and then radiated energy. It can be
expressed in terms of an effective temperature \Teff, and the second parameter
is the half surface mass density $\Sigma$ of the disk ring:
$$T_{\rm eff}^4(R)=[1-(R_\star/R)^{1/2}]\,3GM_\star\dot{M}/8\sigma\pi R^3 \qquad
\Sigma(R)=[1-(R_\star/R)^{1/2}]\, \dot{M}/3\pi \bar{w}.$$ $\sigma$ and $G$ are
the Stefan-Boltzmann and gravitational constants, respectively. $\bar{w}$ is the
depth mean of viscosity $w(z)$, where $z$ is the height above the disk
mid-plane. The viscosity is given by the standard $\alpha$-parametrization as a
function of the total (i.e.\ gas plus radiation) pressure, but numerous other
modified versions are used in the literature. We use a formulation involving the
Reynolds number $Re$, as proposed by \citep{kriz86} and we set $Re=15\,000$ which
corresponds to $\alpha\approx 0.01$.

For the results presented here we selected the following parameters. 
The NS mass is 1.4~\msun. The radii of the inner and outer
disk edges are 2000 and 200\,000~km, respectively. The disk is represented by
nine rings or, more precisely, radial grid points. The accretion
rate is  $\dot{M}=3\cdot
10^{-9}$~\msun/yr, corresponding to 30\% of the Eddington rate. 
A detailed discussion of the model characteristics can be found in \citep{werner07}.

The {\bf vertical structure} of each disk ring is determined from the
simultaneous solution of the radiation transfer equations plus the structure
equations. The latter invoke radiative and hydrostatic equilibrium plus charge
conservation and also consist of the NLTE rate equations for the atomic
population densities. The solution of this set of highly non-linear
integro-differential equations is performed using the Accelerated Lambda
Iteration technique \citep{werner03}. Our code allows for the irradiation
of the disk by the central source, however, the results presented here are
computed with zero incident intensity.

The radiation-transfer equations plus vertical structure equations are solved
like in the stellar atmosphere case, but accounting for two basic
differences. First, the gravity (entering the hydrostatic equation for the
total, i.e.\ gas plus radiation, pressure) is not constant with depth, but
increases with $z$. The gravity is the vertical component of the gravitational
acceleration exerted by the central object (self-gravitation of the disk is
negligible). Second, the energy equation for radiative equilibrium balances the
dissipated mechanical energy and the net radiative losses.

For each atomic level $i$ the NLTE rate equation describes the equilibrium of rates
into and rates out of this level and, thus, determine the occupation numbers
$n_i$. The rate coefficients have radiative and electron collisional
components.  The blanketing by millions of lines from iron  arising from
transitions between some $10^5$ levels can only be attacked with the help of
statistical methods \citep{dreizler93}. We have created a detailed iron model
atom for the NLTE calculations. It comprises the lowest eleven ionisation stages
and a total number of more than 3 million lines.

\begin{figure}
\centering \hspace{-3mm}
\includegraphics[width=0.39\textwidth,angle=90]{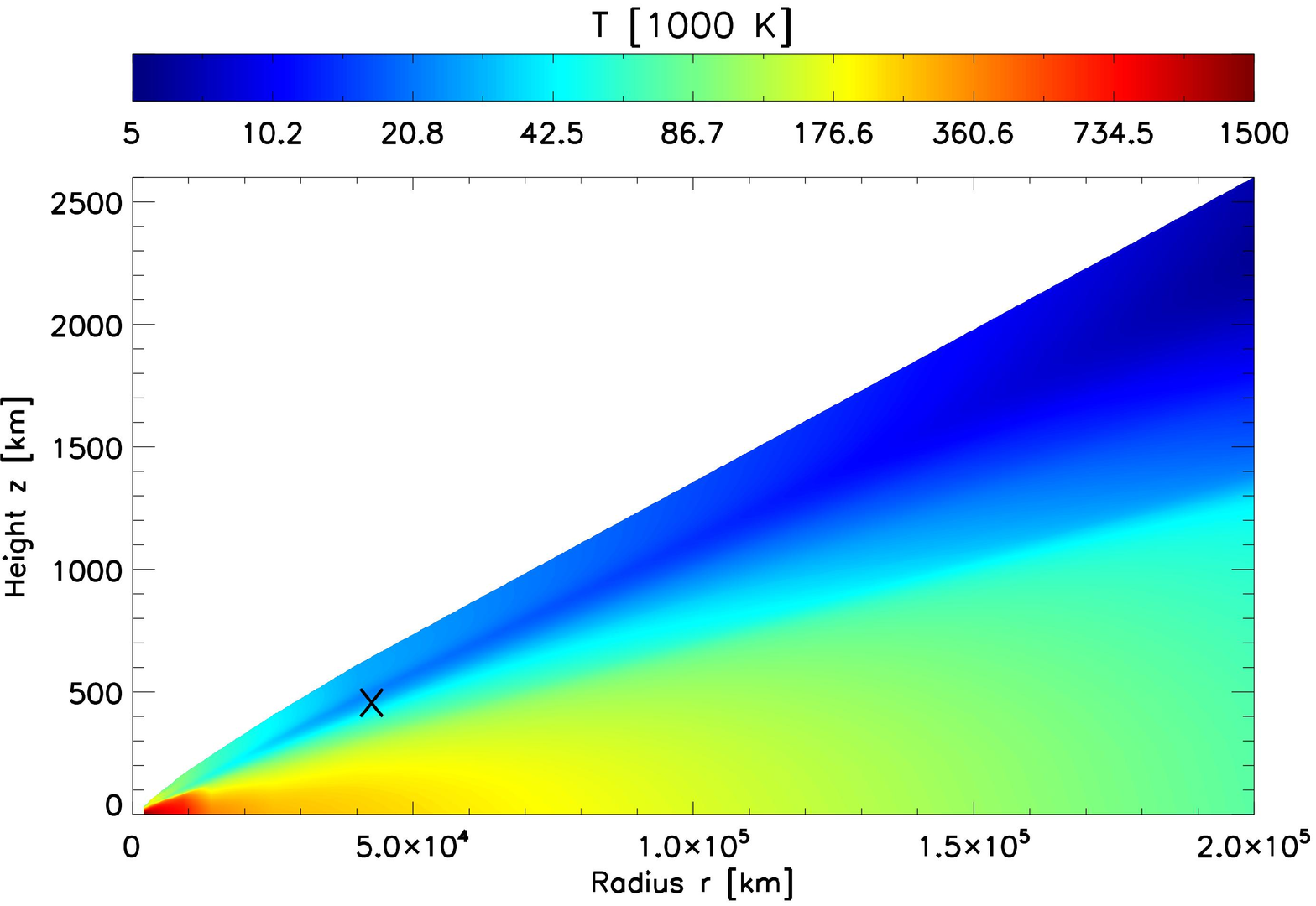}
\includegraphics[width=0.45\textwidth]{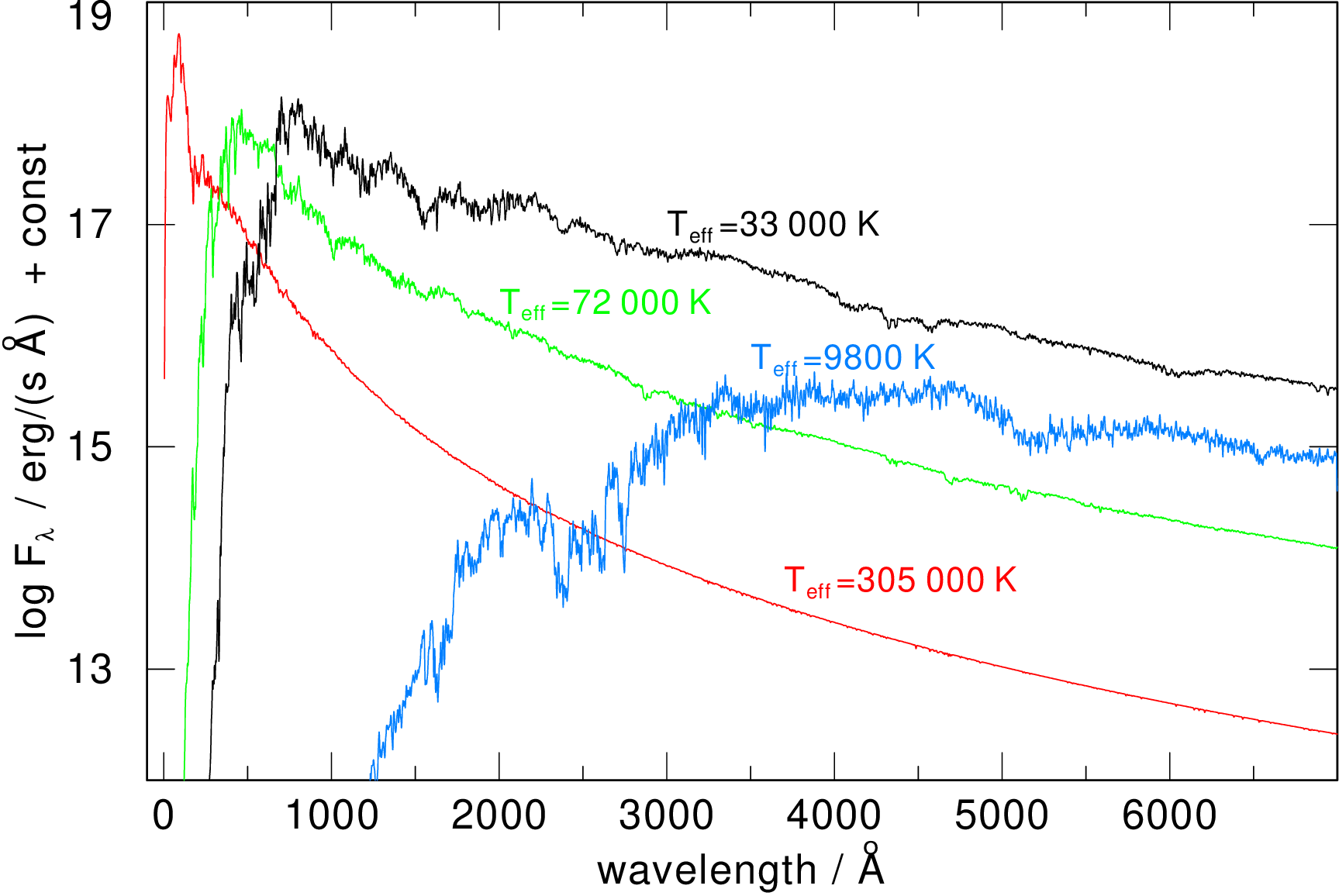}
\caption{\emph{Left:} This cut perpendicular to the midplane shows the
  temperature  structure of the disk. Note that the vertical scale (height above
the midplane) is expanded.  The cross marks the depth at $R$=40\,000~km where
$\tau_{\rm Ross}=1$. \emph{Right:} Relative contribution of four individual disk
rings to the total disk flux. The flux from the ring at $R$=40\,000~km with
\Teff=33\,000~K dominates the total disk spectrum at UV/optical wavelengths.}
\label{fig1}
\end{figure}

The composition of the fallback material in the disk is not
exactly known. It depends on the amount of mass that goes into the disk. A disk
with a small mass ($\lappr 0.001$~\msun) will be composed of Si-burning
ash \citep{menou01}. For simplicity, the results presented here are obtained by
assuming a pure-Fe composition. Tests show that the emergent spectrum is
insensitive against the exact composition as long as Fe is the dominant
species. 

Having calculated the vertical structures and spectra of the individual disk
rings, the ring spectra are integrated taking into account Keplerian rotation.

\section{Results}

The left panel in Fig.~\ref{fig1} displays the temperature structure of the
disk. The temperature varies between 1.5 million K in the midplane at the inner
disk edge down to 6000~K in the upper layers at the outer disk edge.

Which disk regions contribute to the total disk spectrum and to what extent? In
the right panel of Fig.~\ref{fig1} we plot the emergent astrophysical flux from
the area of four disk rings, i.e., the computed flux per cm$^2$ is weighted with
the ring area. The spectral flux distribution of the innermost ring with
\Teff=305\,000~K has its peak value in the soft X-ray region. The contribution
of this disk region to the optical/UV spectrum is negligible. Cutting off
the disk at this inner radius ($R$=2000~km), therefore, is justified if this
spectral range is of interest. The disk region that is dominating the UV/optical
flux is represented by the ring at $R$=40\,000~km with \Teff=33\,000~K.
Its spectrum is dominated by strong blends of the numerous iron lines. Further
out in the disk \Teff\ decreases and the flux contribution to
the UV/optical spectrum declines, too. Our outermost ring has
\Teff=9800~K, its flux maximum is at $\lambda$=4000~\AA\ and it is fainter than
the inner neighbor ring over the whole spectral range. Cutting off the disk at
this outer radius ($R$=200\,000~km) therefore does not affect the UV/optical
spectral region.

\begin{figure}
\includegraphics[width=0.6\textwidth]{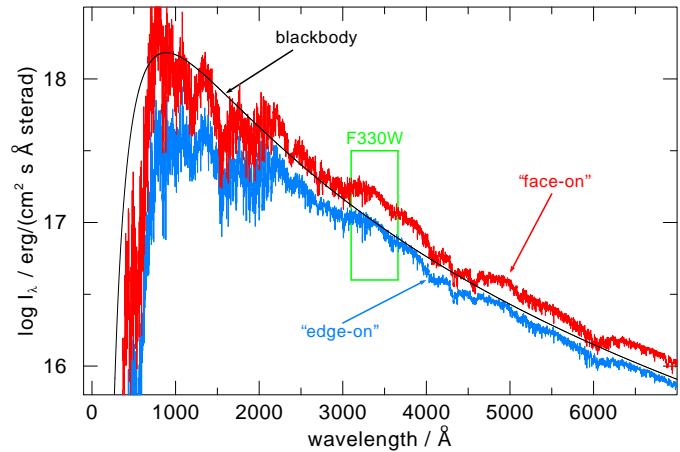} 
\caption{Effect of limb darkening: Specific intensity of the disk at
  $R$=40\,000~km seen under
  inclination angles $87^\circ$ (i.e. almost edge-on) and
  $18^\circ$ (i.e. almost face-on). For comparison we also show a
  blackbody spectrum with $T$=\Teff$(R)=33\,000$~K. The rectangle indicates the
  passband of the HST/ACS filter F330W: The blackbody spectrum systematically
  underestimates the flux for intermediate inclinations.} \label{fig2}
\end{figure}

Our model spectra show distinct limb-darkening effects. The situation is similar
to the stellar atmosphere case (center-to-limb variation of the specific
intensity). Looking face-on we see into deeper and hotter (and thus
``brighter'') layers of the disk when compared to a more edge-on view. In
Fig.~\ref{fig2} we compare the specific intensity emitted at $R$=40\,000~km (per unit
area) for a high and a low inclination angle. Overall, the ``edge-on'' spectrum
is roughly a factor of two fainter than the ``face-on'' spectrum in the optical
region. The difference increases towards the UV. 
We conclude that limb darkening effects are important when disk
dimensions are to be estimated from magnitude measurements.

We now compare the intensities with a blackbody spectrum
(Fig.~\ref{fig2}). Depending on the wavelength band, the blackbody over- or
underestimates the ``real'' spectrum up to a factor of two in the optical and a
factor of four in the UV. The upper limit of $R=10^5$~km for the \sn\ disk
extension at this particular accretion rate was derived from an HST observation
performed with the F330W filter. Its passband is indicated in
Fig.\ref{fig2}. Assuming a disk inclination that coincides with that of the
observed inner equatorial ring around \sn\ ($i=43^\circ$, \citep{panagia91}) one
can see that the blackbody assumption systematically underestimates the flux in
this wavelength region by about 20\%. This means that the observed flux limit
together with our model imposes an even stronger limit on the disk extension. In
order to reduce the disk luminosity in this wavelength region by 20\% its
extension must be reduced by about 30\%. Hence we
find that the upper limit for the disk extension must be reduced from $100\,000$~km
to $70\,000$~km.

\vspace{3mm}\noindent
{\bf Acknowledgement} \quad T.R. is supported by the German Ministry of
Education and Research through DESY under grant 05\,AC6VTB.


\begin{thebibliography}{}

\bibitem{alp01} Alpar, M.~A.  ApJ, {\bf 554}, 1245 (2001)
\bibitem{chat00} Chatterjee, P., Hernquist, L., \& Narayan, R. ApJ, {\bf 534},
  373 (2000)
\bibitem{dreizler93} Dreizler, S., \& Werner, K.  A\&A, {\bf 278}, 199 (1993)
\bibitem{ertan07}  Ertan, \"U., Erkut, M.H., Ek{\c s}i, K.~Y., Alpar, M.~A. {\bf ApJ,
  657}, 441 (2007)
\bibitem{ertan04} Ertan, \"U., \& Cheng, K.~S. ApJ, {\bf 605}, 840 (2004)
\bibitem{graves05} Graves, G.~J.~M., Challis, P.~M., Chevalier, R.~A.,
  et~al.  ApJ, {\bf 629}, 944 (2005)
\bibitem{israel04} Israel, G.~L., Rea, N., Mangano, V., et~al. ApJ, {\bf 603},
  L97 (2004)
\bibitem{kern02} Kern, B., \& Martin, C. Nature, {\bf 417}, 527 (2002)
\bibitem{kriz86} Kriz, S., \& Hubeny, I.
  Bull. Astron. Inst. Czech., {\bf 37}, 129 (1986)
\bibitem{menou01} Menou, K., Perna, R., \& Hernquist, L. ApJ, {\bf 559}, 1032 (2001)
\bibitem{meyer92} Meyer-Hofmeister, E.  A\&A, {\bf 253}, 459 (1992)
\bibitem{Na04} Nagel, T., Dreizler, S., Rauch, T., \& Werner, K.  A\&A,
  {\bf 428}, 109 (2004)
\bibitem{panagia91} Panagia, N., Gilmozzi, R., Macchetto, F., Adorf, H.-M., \&
  Kirshner, R.~P. ApJ, {\bf 380}, L23 (1991)
\bibitem{vanp95} van Paradijs, J., Taam, R.~E., \& van den Heuvel,
  E.~P.~J. A\&A, {\bf 299}, L41 (1995)
\bibitem{perna00} Perna, R., Hernquist, L., \& Narayan, R. ApJ, {\bf 541},
  344 (2000)
\bibitem{shakura73} Shakura N.~I., \& Sunyaev R.~A.  A\&A, {\bf 24}, 337 (1973)
\bibitem{wang06} Wang, Z., Chakrabarty, D., \& Kaplan, D.~L. Nature, {\bf 440},
  772 (2006)
\bibitem{werner03} Werner, K., Deetjen, J.~L., Dreizler, S., et~al. in: Hubeny, I.,
        Mihalas, D., Werner, K. (ed.) Stellar Atmosphere Modeling, ASP
        Conf. Series, {\bf 288}, 31 (2003)
\bibitem{werner07} Werner, K., Nagel, T., \& Rauch, T. Ap\&SS, in press, astro-ph/0608529  (2007)
\bibitem{woods06} Woods, P.~M., \& Thompson, C. in: Lewin, W.~H.~G., van der Klis,
  M. (ed.) Compact Stellar X-Ray Sources, Cambridge University Press, p.~547 (2006)

  
\end{thebibliography}
\end{document}